\documentclass[a4paper,12pt]{article}
\input epsf.sty
\begin{document}
\begin{center}
{\Large {\bf The Source of Maser Emission~W33C (G~12.8$\mathbf{-}$0.2)} }
\medskip
 {\large {\bf  P.~Colom$^1$, E.~E.~Lekht$^2$, M.~I.~Pashchenko$^2$,\\
and G.~M.~Rudnitskij$^2$}}
\end{center}

\medskip
{\it $^1$LESIA, Observatoire de Paris, Section de Meudon, 5 place
Jules Janssen, Meudon, 92195 France

$^2$M.~V.~Lomonosov Moscow State University, Sternberg Astronomical Institute,\\
13 Universitetskij prospekt, Moscow, 119234 Russia\\
{\rm e-mail:   lekht@sai.msu.ru}
}

\bigskip

\begin{abstract}
Results of observations of the maser sources toward the W33C
region (G12.8\mbox{-}0.2) carried out on the 22-m radio telescope
of the Pushchino Radio Astronomy Observatory  in the 1.35-cm
H$_2$O line and on the Large radio telescope in Nan{\c{c}}ay
(France) in the main (1665 and 1667~MHz) and satellite (1612 and
1720~MHz) OH lines are reported. Multiple, strongly variable
short-lived H$_2$O emission features were detected in a broad
interval of radial velocities, from $-$7 to 55~km/s. OH maser
emission in the 1667-MHz line was discovered in a velocity range
of 35--41~km/s. Stokes parameters of maser emission in the main
OH lines 1665 and 1667~MHz were measured. Zeeman splitting was
detected in the 1665-MHz line at 33.4 and 39.4~km/s and in the
1667~MHz line only at 39.4~km/s. The magnetic field intensity was
estimated. A appreciable variability of Zeeman splitting
components was observed at 39 and 39.8~km/s in both main lines.
The extended spectrum and fast variability of the H$_2$O maser
emission together with the variability of the Zeeman splitting
components in the main OH lines can be due to the composite
clumpy structure of the molecular cloud and to the presence in it
of large-scale rotation and bipolar outflow as well as of
turbulent motions of material.
\end{abstract}

The maser radio emission in the OH lines toward the HII region in
the source W33C (G12.8$-$0.2) was detected by Pashchenko in 1975
[1,~2], and in the 1.35-cm water vapor line by Genzel and Downes
in 1976 [3].

In this region the radio continuum source has two emission
peaks. Observations at 408~MHz [4] and 5000~MHz [5] showed that
the faintest component, G12.7$-$0.2, has a size of $5'\times
4'$, while the size of the intense component, G12.8$-$0.2, is
only $0.8'$. The kinematic distance to G12.8$-$0.2 was estimated
from the neutral-hydrogen absorption line profile as 5~kpc [6].

In January 1975 and October 1978 we carried out observations
toward the continuum source W33C on the Large radio telescope in
Nan{\c{c}}ay (France) in all four 18-cm OH lines, in both
circular polarizations [1,~2]. Strongly polarized maser emission
at velocities of 32.5 and 34.5~km/s on the background of strong
absorption toward the HII region G12.8$-$0.2 was detected in the
main 1665-MHz OH line. In the 1667-MHz line we observed faint
emission, also on the background of strong absorption. We also
determined the coordinates of the OH emission source:
$\alpha_{1950} =
18^{\mathrm{h}}11^{\mathrm{m}}18.5^{\mathrm{s}}$, $\delta_{1950}
= -17^{\circ}56'\pm 1.5'$. The OH maser is embedded in a compact
molecular cloud, which is a source of type~IIc OH emission in
the 1612- and 1720-MHz satellite lines.

Subsequently, we conducted observations of the OH maser W33C in
1991 and 2008--2011. Unfortunately, in the literature there is
no information about observations of this OH maser by other
authors, either on single antennas or on high angular resolution
systems. It should be noted that the W33 region hosts two
centers of the 1665-MHz maser emission arranged symmetrically at
both sides from W33C.

In November 1976 Genzel and Downes [3] detected on the 100-m
radio telescope in Effelsberg intense H$_2$O maser emission of
the source W33C at a wavelength of 1.35~cm in an interval of
radial velocities from $-$4 to 1~km/s (8~Jy) and weaker emission
(less than 4~Jy) at velocities from 30 to 41~km/s. The
coordinates of the detected source to within measurement errors
coincided with those of the OH maser source. Later, the H$_2$O
maser W33C was observed by Jaffe et~al. [7] and Comoretto et~al.
[8]. In 1981 the emission was notably weaker and took place at
$-$7 and 34~km/s.

\section{OBSERVATIONS AND DATA}

The observations of the 1.35-cm H$_2$O maser radio emission in a
line 1.35~cm toward the source~W33C ($\alpha_{1950} =
18^{\mathrm{h}}11^{\mathrm{m}}18.3^{\mathrm{s}}$, $\delta_{1950}
= -17^{\circ}56^{\prime}21^{\prime\prime}$) were carried out on
the 22-m RT-22 radio telescope (Pushchino) in November 1981, and
then since March 2010 to November 2011. The noise temperature of
the system with a cooled frontend FET amplifier in the
observations of this source was 120--270~K depending on weather
conditions.

The signal analysis was implemented by a 2048-channel
autocorrelator with a spectral resolution of 6.1~kHz
(0.0822~km/s at 22~GHz). For a pointlike source an antenna
temperature of 1~K corresponds to a flux density of 25~Jy [9].

The observations of the radio source~W33C in the 18-cm hydroxyl
lines were conducted on the radio telescope of the Nan{\c{c}}ay
Radio Astronomy Station of the Paris--Meudon Observatory
(France) at different epochs. The telescope is a Kraus system
two-mirror instrument making possible observations of radio
sources near the meridian. Using a spherical mirror allows us,
by moving the feed, to track a radio source within $\pm
30^\mathrm{m}/\cos\delta$ in the hour angle with respect to the
meridian. At declination $\delta = 0^{\circ}$ the telescope
beamwidth at a wavelength of 18~cm is $3.5'\times 19'$ in the
right ascension and declination, respectively. The telescope
sensitivity at $\lambda = 18$~cm and $\delta = 0^{\circ}$ is
1.4~K/Jy. The noise temperature of the helium-cooled amplifiers
is from 35 to 60~K, depending on conditions of the observation.

The spectral analysis was conducted by an autocorrelation
spectrum analyzer, the frequency resolution was 763~Hz. In the
1665- and 1667-MHz lines this corresponds to a radial-velocity
resolution of 0.137~km/s. In the observations of 2010--2011 the
resolution was twice as high, 0.068~km/s. Since 2008 the radio
telescope simultaneously receives two perpendicular modes of
linear polarization, which yield directly the intensities of the
corresponding linear modes (L~$0^{\circ}$, L~$90^{\circ}$).
Mixing signals from the perpendicular feeds with a phase delay of
one of the modes by a quarter of the wavelength produces two
orthogonal circular modes (LC, RC). Thus, three Stokes parameters
$I$, $V$, and $Q$ are in fact observed simultaneously (with an
appropriate choice of the coordinate system).

The observations were processed by the GILDAS software package
of the Institute of Millimeter Radio Astronomy (IRAM, Grenoble,
France), available on the Web at
\texttt{http://www.iram.fr/IRAMFR/GILDAS/} [9].

\begin{figure}[t!]
\centering\leavevmode \epsfysize=17cm \epsfbox{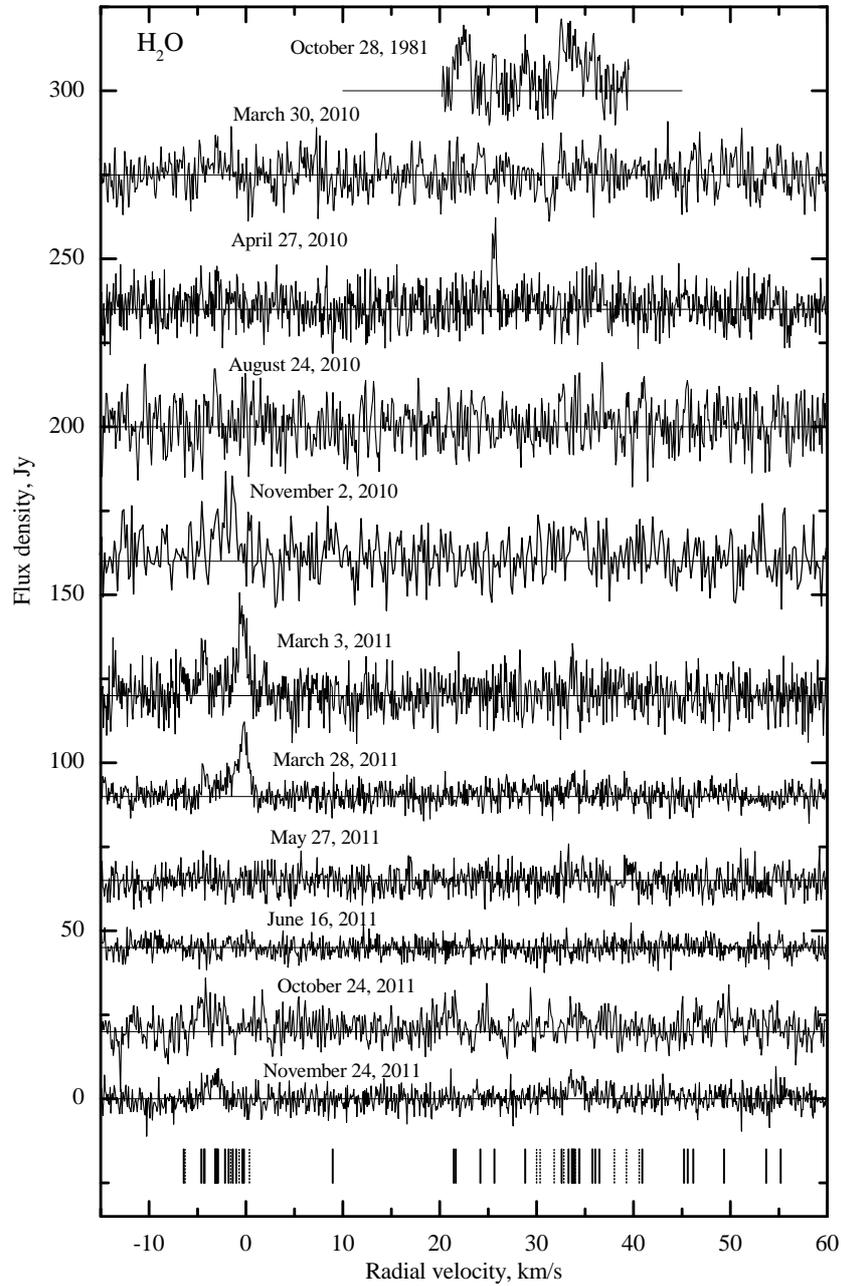}
\caption{Spectra of the H$_2$O maser emission toward the
source~W33C. A double-pointed arrow shows the scale. The radial
velocity is given with respect to the Local Standard of Rest.
Vertical bars at the bottom mark the velocities at which
emission features were ever observed by us (solid curves) and
other authors (dashed curves).}
\label{fig1}
\end{figure}

Figure~\ref{fig1} presents the H$_2$O spectra for different
epochs. A double-pointed arrow shows the scale in janskys. The
horizontal axis is the velocity with respect to the Local
Standard of Rest. Vertical bars at the bottom mark the velocities
at which emission features were ever observed either by us or by
other authors [3,~7,~8].

\begin{figure}[t!]
\centering\leavevmode \epsfysize=8cm \epsfbox{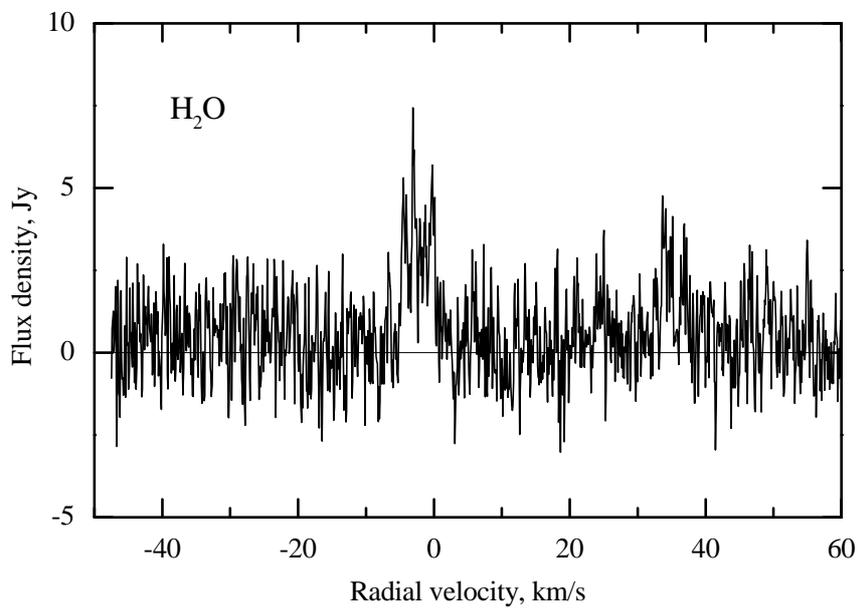}
\caption{Averaged spectrum of the H$_2$O maser emission in~W33C
for a time interval of 2010--2011.}
\label{fig2}
\end{figure}

The H$_2$O maser emission was mostly observed in two spectral
intervals: from $-$7 to $+$1~km/s and from 32 to 37~km/s
(Fig.~\ref{fig2}). Single emission features scattered in the
spectrum from 9 to 55~km/s were also observed. Thus, the full
velocity interval is about 62~km/s.

\begin{figure}[t!]
\centering\leavevmode \epsfysize=18cm \epsfbox{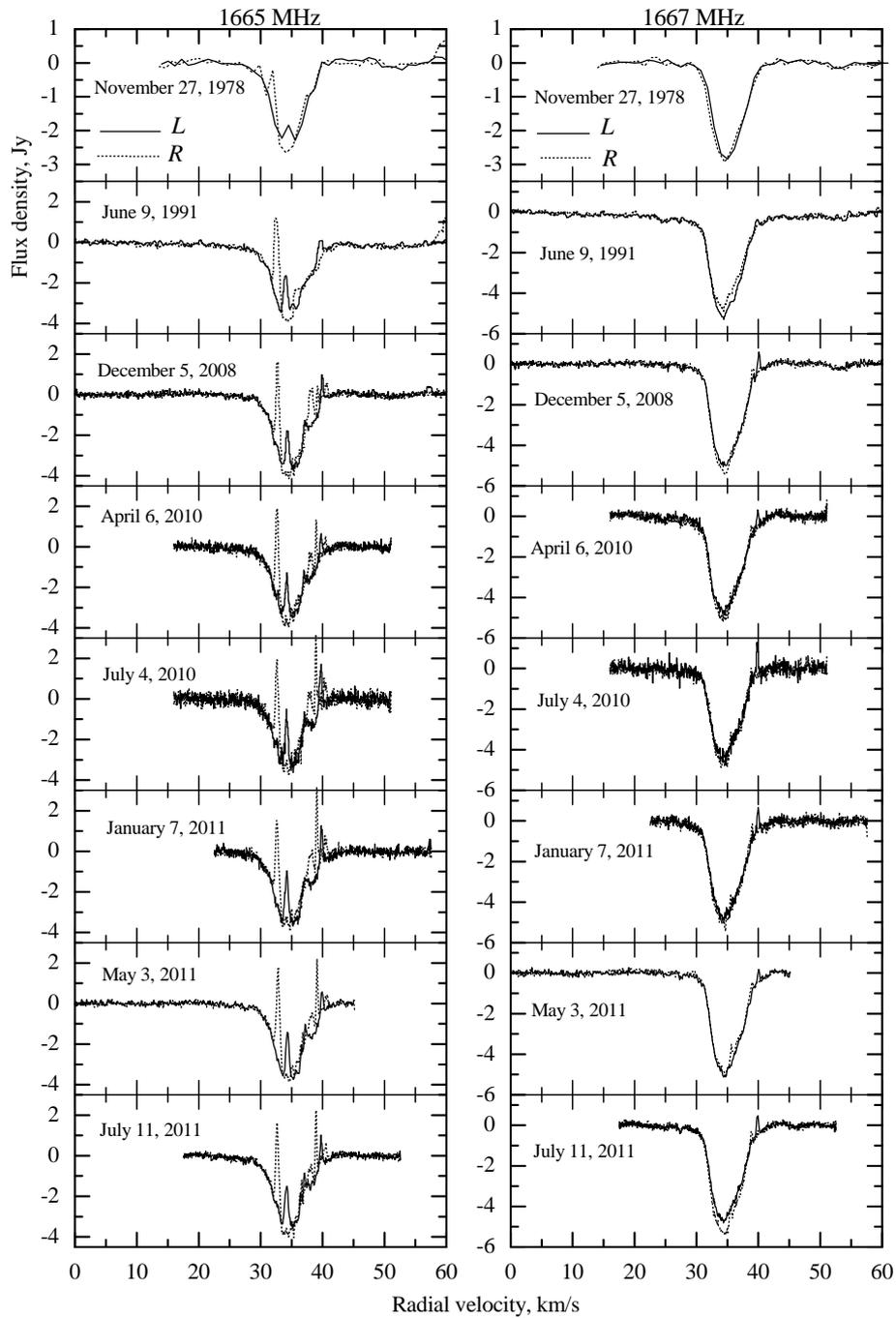}
\caption{Spectra of maser emission in the 1665- and 1667-MHz
hydroxyl lines for left-hand~(L) and right-hand (R) circular
polarizations at different epochs of the observations.}
\label{fig3}
\end{figure}

The results of our observations of the hydroxyl maser emission in
the 1665- and 1667-MHz lines at different epochs are shown in
Fig.~\ref{fig3}. In 1975 and 1978 the observations were carried
out with a resolution of 1~km/s, in 2008 with a resolution of
0.137~km/s, and in 2010--2011 with a resolution of 0.068~km/s.
Solid curves show emission in the left-hand and dashed ones in
the right-hand circular polarizations. The technique of the
observations is described in [10,~11].

The results of the observations in the satellite lines (1612 and
1720~MHz) are shown in Fig.~\ref{fig4}.

For the main lines 1665 and 1667~MHz Figures~\ref{fig5} and
\ref{fig6} show Stokes parameters for the epochs of December 5,
2008, and May 3, 2011, respectively. Time variations of Stokes
parameter $V$ for the central part of the 1665-MHz spectrum are
presented in Fig.~\ref{fig7}, and for the 1667-MHz line in
Fig.~\ref{fig8}. The main components are numbered, and their
parameters are listed in Tables~\ref{tab1} and \ref{tab2}, where
$F_1$, $F_2$, $F_3$, $F_4$ are flux densities of the components,
$\delta V_{2,1}$ and $\delta V_{4,3}$ are velocity differences
between components~\emph{2} and~\emph{1} and components~\emph{4}
and~\emph{3}, respectively.

\section{DISCUSSION OF THE RESULTS}

Though the maser emission source W33C was discovered long ago,
it has been poorly studied either in the water-vapor line or in
the hydroxyl lines. High angular resolution observations lack,
and this does not allow us to spatially localize the observed
emission features. Therefore, in the data analysis we
concentrate on the variability of the maser emission.

\subsection{H$_2$O Maser Emission}

In spite of the fact that regular observations of the H$_2$O
maser emission in G12.8$-$0.2 were conducted by us only since
the beginning of 2010, we managed to find a number of important
peculiarities of this emission. Let us list the main ones.

1.~The emission spectrum is rather broad, from $-7$ to 55~km/s.

2.~Most emission features are short-lived.

3.~There are two groups of persistent emission features with a
difference in radial velocities of $\sim$37~km/s; the mean
velocity of the second group (34~km/s) coincides with the
central velocity of the OH absorption line and with the velocity
of the OH maser emission.

4.~There is a weak emission at 24~km/s, which was clearly visible
after averaging the H$_2$O spectra (Fig.~\ref{fig2}). This
suggests this emission is faint but rather stable.

In addition to the H$_2$O spectrum, we have also detected a
continuum signal with $T_{\mathrm{a}}\sim 0.5$~K, which
corresponds to a flux density of $\sim$15~Jy.

\subsection{Hydroxyl Emission}

In all OH main-line profiles we observe strong absorption, on
which the maser emission is superimposed. The absorption line
with a central velocity of 35.8~km/s arises in a compact
molecular cloud, which is also a source type~IIc OH emission in
the 1612- and 1720~MHz lines [2]. The absorption linewidth at
half-maximum is 5.8~km/s.

The variability of the OH maser emission in W33C is best
manifested in Stokes parameter $V$; therefore, our subsequent
analysis is mostly connected with variations of this parameter.

The amplitude of features~\emph{1} and~\emph{2} observed only in
the 1665-MHz line varies very weakly, whereas features~\emph{3}
and~\emph{4} undergo considerable amplitude variations in both
main lines (Tables~\ref{tab1} and \ref{tab2}). The listed pairs
of features arise owing to splitting of the emission at 33.4 and
39.4~km/s in a longitudinal magnetic field into two components
shifted in velocity up and down and elliptically polarized in
opposite directions.

In the 1665-MHz line the difference in the radial velocities of
features~\emph{2} and~\emph{1} is 1.7~km/s, and between
features~\emph{4} and~\emph{3} it is 0.8~km/s. A splitting of
1.7~km/s in the emission at 33.4~km/s corresponds to a
longitudinal magnetic field of 2.9~mG, and a splitting of
0.8~km/s at 39.4~km/s to 1.4~mG.

In the emission detected by us in the 1667~MHz line, the
separation between the Zeeman components \emph{4} and~\emph{3} is
$\approx 1$~km/s (Table~\ref{tab2}); this corresponds to a
magnetic field of 2.8~mG. The twofold difference in the magnetic
field intensity for the emission at 39.4~km/s found from the
splitting in the main OH lines can be due to various causes. The
emission regions can be close however not coinciding or have
different sizes (the region radiating in the 1667-MHz line is more
extended). In~both cases the medium should be inhomogeneous,
turbulent.

\begin{figure}[t!]
\centering\leavevmode \epsfysize=13cm \epsfbox{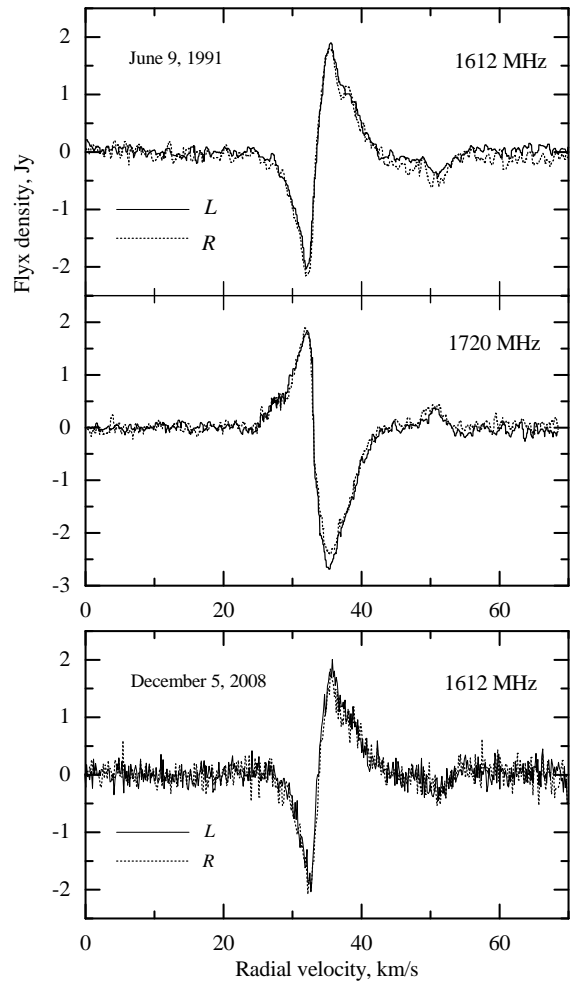}
\caption{Same as in Fig.~3, but for the 1612 and 1720-MHz
satellite lines.}
\label{fig4}
\end{figure}

\begin{figure}[t!]
\centering\leavevmode \epsfysize=13cm \epsfbox{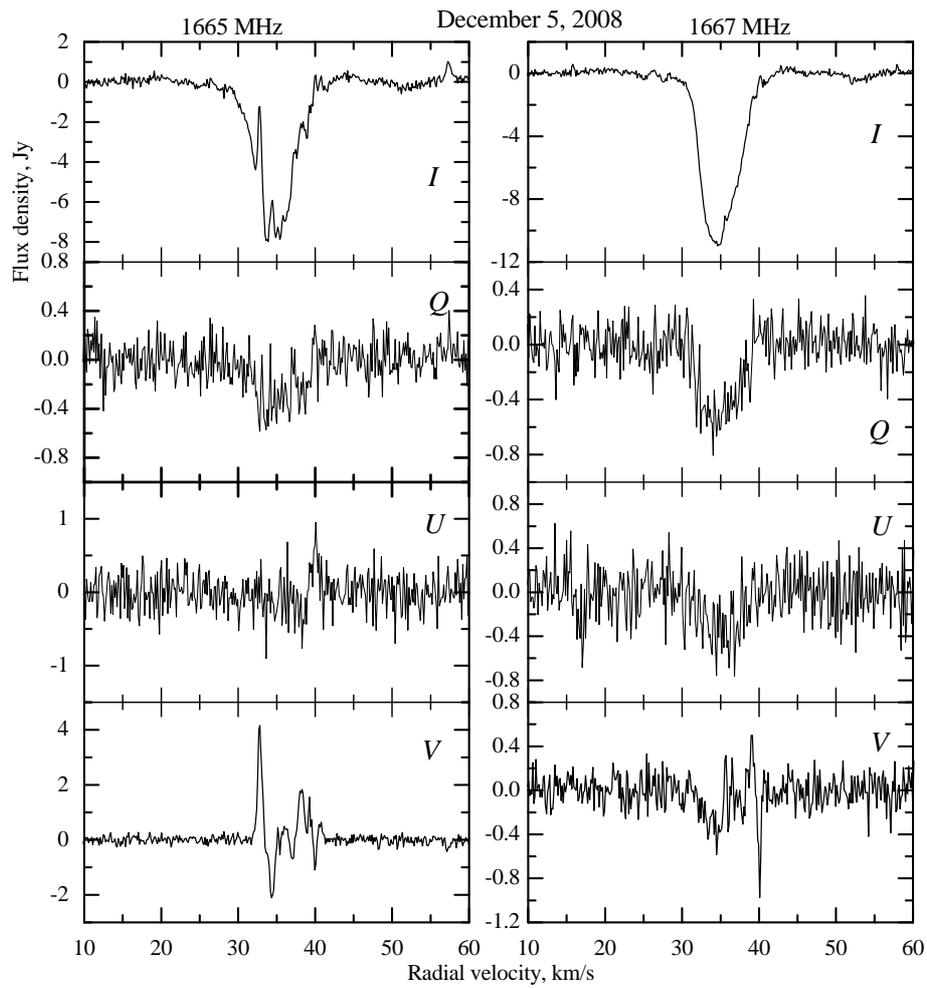}
\caption{Stokes parameters for the emission in the 1665- and
1667-MHz hydroxyl lines at the epoch of December 5, 2008.}
\label{fig5}
\end{figure}

\begin{figure}[t!]
\centering\leavevmode \epsfysize=13cm \epsfbox{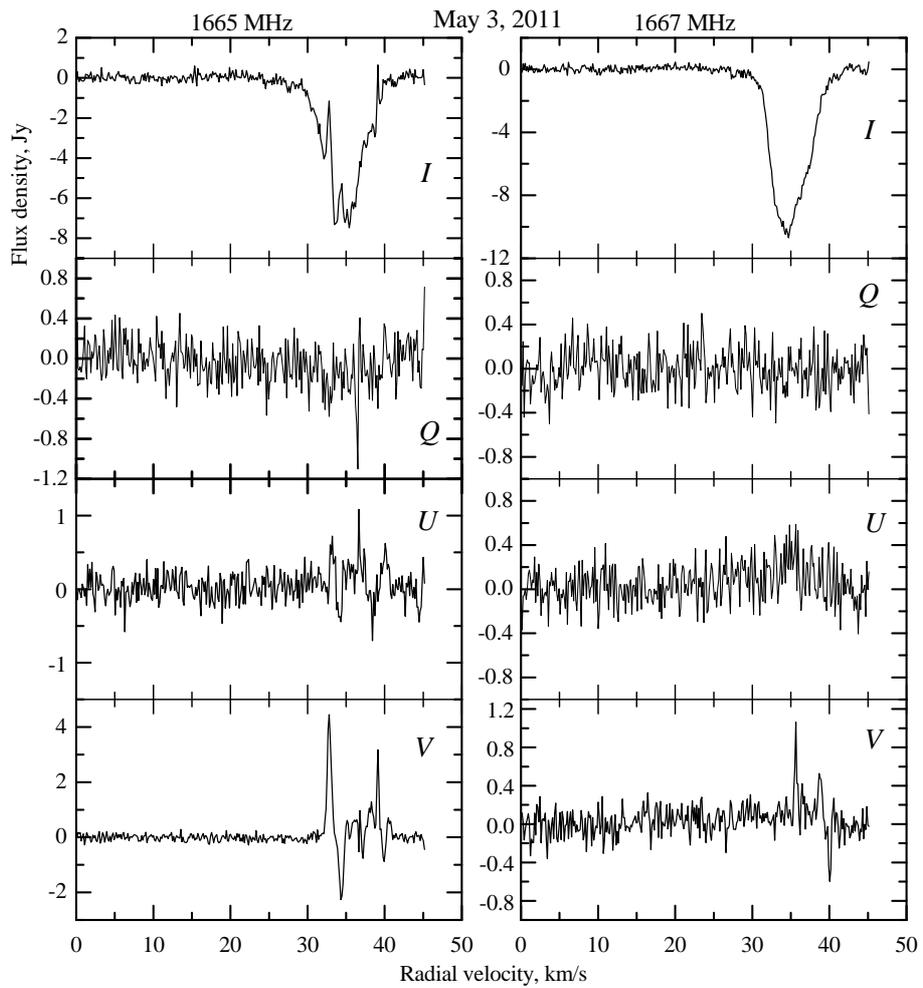}
\caption{Same as in Fig.~5, but for the epoch of May 3, 2011.}
\label{fig6}
\end{figure}

\begin{figure}[t!]
\centering\leavevmode \epsfysize=14cm \epsfbox{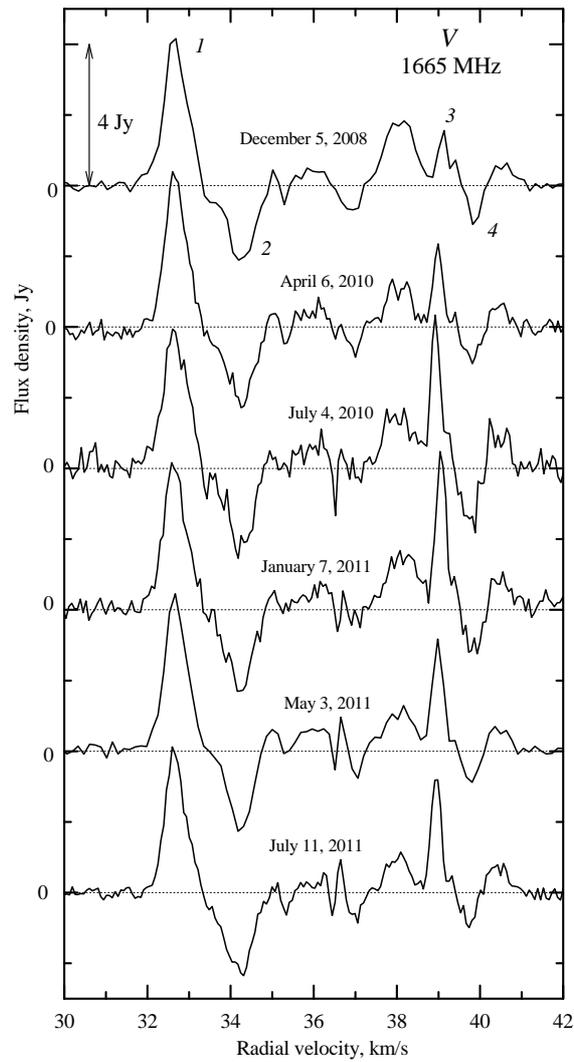}
\caption{Time variations of Stokes parameter $V$ for the
central part of the 1665-MHz spectrum. The main features are
numbered.}
\label{fig7}
\end{figure}

\begin{figure}[t!]
\centering\leavevmode \epsfysize=8cm \epsfbox{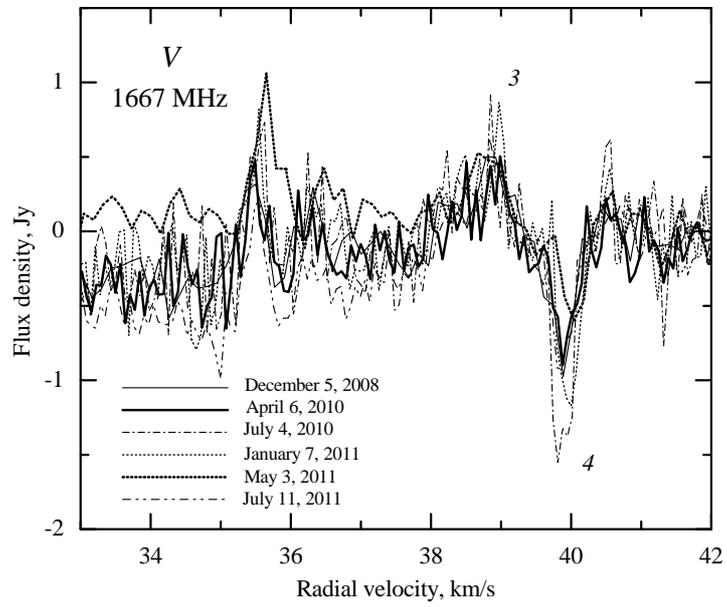}
\caption{Same as in Fig.~7, but for the central part of the
1667-MHz spectrum.}
\label{fig8}
\end{figure}

The profiles of the 1612- and 1720-MHz OH lines (Fig.~\ref{fig4})
consist of emission and absorption components and mirror each
other. Such a structure finds its explanation in the model of an
OH source that is associated with a molecular cloud around the
maser in the presence of an embedded IR emission source affecting
the populations of hyperfine-structure sublevels of OH molecules
[12]. The particulars of pumping by IR emission in model [12] are
such that the inversion in the 1720-MHz transition is accompanied
by antiinversion in the 1612-MHz line, and vice versa.
Observations of a number of sources in the satellite lines of OH
[13,~14] indeed demonstrate mirror profiles of the lines:
1612~MHz in emission, 1720~MHz in absorption, and vice versa. If
a magnetic field is present in the cloud, then, according to
model [12], inversion or antiinversion of this or that OH
satellite transition is determined by the angle between the
direction of propagation of IR emission and the local field
vector. With the source inside the cloud, in some parts of it
inversion in one of the satellites and antiinversion in the other
will be observed; in other parts of the cloud the situation is
opposite. If the cloud is not resolved by the radio telescope
beam, the superposition of profiles of satellite lines from
different parts of the cloud produces a pattern similar to that
in Fig.~\ref{fig4}.

\subsection{Model of the Maser Source in~W33C}

Let us summarize the results of our OH and H$_2$O observations.
Common for them is that the hydroxyl emission and emission of a
group of water-vapor features take place at the same radial
velocities (30--40~km/s). Note that the emission in other
molecular lines (e.g., CH$_3$OH [15]) and in radio recombination
lines (e.g., H134$\alpha$ [16]) is also observed in this
velocity range. Most likely, these emissions are associated with
the same molecular cloud at a radial velocity of about 35~km/s.

We also note that the profiles of the hydroxyl and methanol
lines have a double-peaked structure; its components are
arranged more or less symmetrically with respect to the velocity
35~km/s. Such a structure of the spectra is best explained by
the model of a rotating molecular cloud. The rotation rate is
low, i.e., the line-of-sight projection of the cloud rotation
rate is small.

The H$_2$O line profiles have a different structure
(Fig.~\ref{fig1} and~\ref{fig2}). A sufficiently stable emission
at velocities from $-7$ to 1~km/s and at a velocity of 24~km/s
can be associated with the bipolar outflow.

\section{MAIN RESULTS}

Let us list the main results of our observations of water-vapor
and hydroxyl masers in the source~W33C.

1.~We have detected a large number of strongly variable,
short-lived H$_2$O emission features in a broad interval of
radial velocities, from $-$7 to 55~km/s. The velocity of one of
the groups with the most stable emission (34~km/s) is close to
the central velocity of the OH absorption line and OH maser
emission.

2.~We have discovered the maser emission of W33C in the 1667-MHz
line in both circular polarizations in the velocity interval
35--41~km/s.

3.~We have found Zeeman splitting in the OH 1665-MHz main line
for the emission at 33.4 and 39.4~km/s. From the splitting
magnitude we have estimated the intensity of the line-of-sight
component of the magnetic field for each of the regions masering
at 33.4 and 39.4~km/s (2.8 and 1.4~mG).

4.~``Mirror'' profiles of the OH satellite lines, 1612 and
1720~MHz, suggest pumping of the levels of corresponding
transitions by IR emission of a source embedded in a magnetized
molecular cloud around the maser.

5.~We have observed an appreciable variability of the 39.0- and
39.8-km/s components of Zeeman splitting in both main lines; at
the same time, the 32.6- and 34.3-km/s components in the
1665~MHz line remained rather stable, and in the 1667-MHz line
they were not detected at all.

6.~The extended spectrum and fast variability of the H$_2$O
maser emission together with the variability of the Zeeman
splitting components at 39~km/s in the main OH lines can be a
consequence of composite clumpy structure of the molecular cloud
and of the presence in it of large-scale (rotation, bipolar
outflow) and turbulent motions of material.

\section*{ACKNOWLEDGMENTS}

This research was supported by the Ministry of Science and
Education of the Russian Federation on the facility RT-22 radio
telescope (registration number~01-10) and by the Russian
Foundation for Basic Research (project code 09-02-00963-a). The
authors are grateful to the staff of the Pushchino (Russia) and
Nan\c{c}ay (France) radio astronomy observatories for the help
with the observations.

\newpage

\begin{table}[t!]
\caption{Main components of the $V$ Stokes
parameter in the 1665-MHz OH line}
\label{tab1}
\def\arraystretch{1.1}
\bigskip
{\footnotesize
\begin{tabular}{|c|c|c|c|c|c|c|c|c|}
\hline Date & $F_1$, Jy & $F_2$, Jy & $|F_1/F_2|$ & $\delta
V_{2,1}$, km/s &
$F_3$, Jy & $F_4$, Jy & $|F_3/F_4|$ & $\delta V_{4,3}$, km/s \\
\hline
2008.12.05 & 4.07 & $-$2.07 & 1.97 & 1.62 & 1.4 & $-$1.0 & 1.4 & 0.69 \\
2010.04.06 & 4.30 & $-$2.2 & 1.95 & 1.63 &2.3 & $-$0.9 & 2.56 & 0.80 \\
2010.07.04 & 3.8 & $-$2.2 & 1.73 & 1.67 & 4.3 & $-$1.4 & 3.07 & 0.88 \\
2011.01.07 & 4.1 & $-$2.3 & 1.78 & 1.62 & 4.3 & $-$2.1 & 2.05 & 0.77 \\
2011.05.03 & 4.3 & $-$2.2 & 1.95 & 1.60 & 3.0 & $-$0.8 & 3.75 & 0.81 \\
2011.07.11 & 4.0 & $-$2.2 & 1.82 & 1.62 & 3.1 & $-$0.9 & 3.44 & 0.79 \\
\hline
\end{tabular}
}
\end{table}
\bigskip

\begin{center}
\begin{table}[t!]
\caption{Main components of the $V$ Stokes
parameter in the 1667-MHz OH line}
\label{tab2}
\bigskip
\centering
\def\arraystretch{1.1}
{\footnotesize
\begin{tabular}{|c|c|c|c|c|}
\hline
Date & F$_3$, Jy & F$_4$, Jy & $|F_3/F_4|$ & $\delta V_{4,3} $, km/s \\
\hline
2008.12.05 & 0.50 & $-$0.85 & 0.59 & 1.02 \\
2010.04.06 & 0.41 & $-$0.76 & 0.54 & 0.99 \\
2010.07.04 & 0.81 & $-$1.38 & 0.59 & 1.02 \\
2011.01.07 & 0.79 & $-$1.15 & 0.69 & 0.98 \\
2011.05.03 & 0.48 & $-$0.57 & 0.84 & 1.23 \\
2011.07.11 & 0.52 & $-$0.53 & 0.65 & 0.96 \\
\hline
\end{tabular}
}
\end{table}
\end{center}

\begin{thebibliography}{99}
\bibitem{1:AstRus1210003Colom}
M.~I.~Pashchenko, Astron. Tsirkulyar No.~886, 1 (1975).
\bibitem{2:AstRus1210003Colom}
M.~I.~Pashchenko, Soviet Astron. Lett. \textbf{6}, 58 (1980).
\bibitem{3:AstRus1210003Colom}
R.~Genzel and D.~Downes, Astron. and Astrophys. Suppl. Ser.
\textbf{30}, 145 (1977).
\bibitem{4:AstRus1210003Colom}
P.~A.~Shaver and W.~M.~Goss, Austral. J.~Phys. Astrophys.
Suppl. No.~14, 77 (1970).
\bibitem{5:AstRus1210003Colom}
W.~M.~Goss and P.~A.~Shaver, Austral. J.~Phys. Astrophys. Suppl.
No.~14, 1 (1970).
\bibitem{6:AstRus1210003Colom}
V.~Radhakrishnan, W.~M.~Goss, J.~D.~Murray, and J.~W.~Brooks,
Astrophys. J. Suppl. Ser. \textbf{24}, 49 (1972).
\bibitem{7:AstRus1210003Colom}
D.~T.~Jaffe, R.~G\"usten, and D.~Downes, Astrophys. J.
\textbf{250}, 621 (1981).
\bibitem{8:AstRus1210003Colom}
G.~Comoretto, F.~Palagi, R.~Cesaroni, M.~Felli, A.~Bettarini,
M.~Catarzi, G.~P.~Curioni, P.~Curioni, S.~di~Franco,
C.~Giovanardi, M.~Massi, F.~Palla, D.~Panella, E.~Rossi,
N.~Speroni, and G.~Tofani, Astron. and Astrophys. Suppl. Ser.
\textbf{84}, 179 (1990).
\bibitem{9:AstRus1210003Colom}
M.~I.~Pashchenko and E.~E.~Lekht, Astron. Reports \textbf{49},
624 (2005).
\bibitem{10:AstRus1210003Colom}
V.~I.~Slysh, M.~I.~Pashchenko, G.~M.~Rudnitski{\u\i},
V.~M.~Vitrishchak, and P.~Colom, Astron. Reports \textbf{54}, 599
(2010).
\bibitem{11:AstRus1210003Colom}
M.~I.~Pashchenko, G.~M.~Rudnitski{\u\i}, and P.~Colom, Astron.
Reports \textbf{53}, 541 (2009).
\bibitem{13:AstRus1210003Colom}
V.~V.~Burdyuzha and D.~A.~Varshalovich, Soviet Astron.
\textbf{16}, 597 (1973).
\bibitem{14:AstRus1210003Colom}
M.~I.~Pashchenko, G.~M.~Rudnitski{\u\i}, and O.~Franquelin,
Soviet Astron. Lett. \textbf{5}, 276 (1979).
\bibitem{15:AstRus1210003Colom}
R.~F.~Haynes and J.~L.~Caswell, Monthly Not. Roy. Astron. Soc.
\textbf{178}, 219 (1977).
\bibitem{16:AstRus1210003Colom}
A.~D.~Haschick, K.~M.~Menten, and W.~A.~Baan, Astrophys. J.
\textbf{354}, 556 (1990).
\bibitem{17:AstRus1210003Colom}
F.~F.~Gardner, T.~L.~Wilson, and P.~Thomasson, Astrophys. Lett.
\textbf{16}, 29 (1975).
\end{thebibliography}
\end{document}